\documentclass[global,onecolumn]{svjour}
\usepackage{graphics}
\usepackage{amsmath}
\usepackage{bbold}
\usepackage{comment}
\journalname{Applied Physics B}
\begin{document}
\title{Double-electron ionization driven by inhomogeneous fields}

\author{A. Chac\'on\inst{1}\thanks{alexis.chacon@icfo.eu} \and L. Ortmann\inst{2} \and F. Cucchietti\inst{3} \and N. Su\'arez\inst{1} \and J. A. P\'erez-Hern\'andez\inst{4} \and M. F. Ciappina\inst{5} \and A. S. Landsman\inst{2,6} \and M. Lewenstein\inst{1,7}
}                     


\institute{ICFO-Institut de Ci\`encies Fot\`oniques, The Barcelona Institute of Science and Technology, Av. Carl Friedrich Gauss 3, 08860 Castelldefels (Barcelona), Spain \and Max Planck Institute for the Physics of Complex Systems, N\"othnitzer Stra{\ss}e 38, D-01187 Dresden, Germany \and Barcelona
Supercomputing Center BSC-CNS, Campus Nord UPC, Barcelona, Spain \and Centro de L\'aseres Pulsados (CLPU), Parque Cient\'{\i}co, E-37185 Villamayor, Salamanca, Spain \and Institute of Physics of the ASCR, ELI-Beamlines, Na Slovance 2, 182 21 Prague, Czech Republic \and Max Planck Postech/Department of Physics, Pohang, Gyeongbuk 37673, Republic of Korea \and ICREA, Pg. Llu\'{\i}s Companys 23, 08010 Barcelona, Spain}

\date{Received: date / Revised version: date}

\maketitle
\begin{abstract}
Electron-electron correlation effects play an instrumental role 
in our understanding of sequential (SDI) and non-sequential 
double ionization (NSDI) mechanisms. Here, we present a 
theoretical study of NSDI driven by plasmonic-enhanced
spatial inhomogeneous fields. By numerically solving the 
time-dependent Schr\"odinger equation for a linear reduced model of 
He and a double-electron time-evolution probability analysis, we provide 
evidence for the enhancement effects in NSDI showing that the double ionization 
yield at lower laser peak intensities is increased due to the inhomogeneity 
of the laser field. Furthermore, our quantum mechanical model, as well as 
classical trajectory Monte Carlo simulations, show that inhomogeneous 
fields are a useful tool for splitting the binary and recoil processes 
in the rescattering scenario.  
\end{abstract}
\section{Introduction}\label{intro}
Since 1982, when L'Huillier presented the first experimental 
observation of a large enhancement in the double-charge 
ionization yield of Xe driven by an intense infrared (IR) laser-field, a number of questions about electron-electron (e-e) correlation
effects and their mechanisms have arisen~\cite{HuillierPRL1982,HuillierPRA1988}. 
The fact that those results could not be explained in the framework of SDI, where e-e correlation effects are assumed negligible, 
opened the path of considering the importance of such correlation effects in the ionization 
processes~\cite{HuillierPRL1982,HuillierPRA1988,BerguesIEEE2015,BerguesNatComm2012,Pullen2016}.
It was then that the concept of Non-Sequential Double Ionization (NSDI)
arose as an explanation of the 1982 experiment ~\cite{BWalkerPRL1994,WeberPRL2000,WeberNat2000,WBeckerRMP2012}. 
However, in the NSDI mechanism there are several processes  
such as the shake-off, laser-field-assisted rescattering ionization, 
Rescattering Impact Double Ionization  (RIDI)~\cite{CorkumPRL1993,CRuizPRL2008,KopoldPRL2000,EreminaJPB2003} and
Rescattering Excitation with Subsequent Ionization (RESI), 
which might take place during the DI of atoms. The question of how to disentangle RIDI and RESI (and within RIDI the binary and recoil processes) are therefore still under investigation in the attosecond science
community~\cite{BerguesIEEE2015}. 


Prior studies addressing e-e correlation effects in laser-driven multiple ionization processes were done considering only spatially homogeneous fields, i.e.~fields that do not present spatial variations in the region where the electron dynamics takes place.
This is a legitimate assumption considering that in conventional laser-matter experiments the laser electric field changes in a region on the orders of micrometers, whereas the electron dynamics develops on a nanometric scale (see e.g.~\cite{reviewROP} for more details). However, since recent studies of post-ionization dynamics in spatially inhomogeneous fields~\cite{AlexisHHG2e} provides new physical effects and insights, a question arises as to the influence of spatial variation on the DI process.  The aim of this work is to present a complete study of DI driven by plasmonic-enhanced spatially inhomogeneous fields with an investigation of NSDI in general, and the RESI and RIDI mechanisms in particular. To this end, we will employ both quantum mechanical approaches, based on the numerical solution of the time-dependent of the Schr\"odinger equation (TDSE) for two electrons in reduced dimensions, and classical schemes employing classical trajectories Monte Carlo (CTMC) simulations. Within the quantum framework, we employ a linear model for the helium atom, where
the motion of both electrons is restricted to the direction of the laser polarization. Experience has shown that 1D models qualitatively reproduce strong-field phenomena such as the double-ionization knee structure~\cite{BauerPRA1997,Lappas1998} or above-threshold
ionization~\cite{ASanpera1} and intense-field double ionization mechanisms~\cite{MLein2000}.

The rest of this paper\footnote{This contribution is dedicated to Ted H\"ansch on the occasion of his 75th birthday. Although Prof. H\"ansch is mostly regarded for "contributions to the development of laser-based precision spectroscopy, including the optical frequency comb technique", and his contributions to laser cooling and physics of ultracold atoms, his influence on attosecond physics is hard to underestimate. For instance, he predicted at very early stages the possibility of generating attosecond pulse trains from phase locked harmonics~\cite{HarrisOpt,Wahlstrom} and pioneered and contributed to the initial studies of the high-order harmonics coherence~\cite{Wahlstrom,Zerne}. His group has also developed decisive steps extending the frequency combs toward high frequencies regimes.} is organized as follows. In the next section we present our theoretical tools, namely the TDSE and CTMC for two electrons in reduced dimensions. Then, in Section 3, we show a comparative study between DI driven by conventional laser pulses and DI governed by plasmonic-enhanced spatially inhomogeneous fields.  We put particular emphasis on the two-electron momentum distribution, considering it represents one of the most detailed observables and is fully experimentally accessible. We end up with concluding remarks and a brief outlook.

\section{Numerical model}\label{sec:1}

We study the two-electron dynamics driven by plasmonic-enhanced fields via a fully quantum mechanical 
linear model of the helium atom and the integration 
of Newton's equations in the framework of a classical trajectory 
Monte Carlo (CTMC) method. The {\it ab}-initio quantum 
mechanical calculations allow us to address the whole 
electron-electron (e-e) correlated dynamics by means 
of the numerical solution of the Time-Dependent 
Schr\"odinger Equation (TDSE) similarly to those used by 
Lein~\cite{MLein2000} and Watson~\cite{ASanpera2,ASanpera1}. 
The Hamiltonian of our 2e system reads (atomic units are used
 throughout the paper unless specified otherwise)
\begin{eqnarray}\label{eq:2eham}
H&=& \sum_{j=1}^2 [\frac{p_j^2}{2} + V(z_j) + V_{\rm int}(z_j,t)] + V(z_1,z_2),
\end{eqnarray}
\noindent where $p_j= -i\frac{\partial}{\partial z_j}$ is the momentum 
operator corresponding to the $j$-th electron ($j$-th-e), $j=1,2$. 
$V(z_j) = -\frac{Z}{\sqrt{z_j^2+a}}$ and 
$V(z_1,z_2)=\frac{1}{\sqrt{(z_1-z_2)^2 +b}}$ 
are the attractive potential 
of the interaction of the $j$-th electron with the nucleus of charge $Z$ and the 
repulsive e-e potential, respectively. The potential describing the interaction of the $j$-th electron with 
the spatially dependent laser electric field in length gauge is ~\cite{AChaconVGPRA2015}
\begin{eqnarray}\label{eq:2elaserInhomo}
V_{\rm int}(z_j,t)&=& (z_j+\frac{\beta}{2}z_j^2)E_{\rm h}(t),
\end{eqnarray}
where $\beta$ denotes the inhomogeneity strength 
(see e.g.~\cite{AChaconVGPRA2015,CiappinaPRA2012} 
for more details) of the plasmonic field and 
$E_{\rm h}(t) = E_0\sin^2(\omega_0t/2{\rm N})\sin(\omega_0t+\varphi_0)$ 
is the spatially homogenous or conventional laser electric field. 
Here, $E_0$, $\omega_0$, $\rm{N}$ and $\varphi_0$ 
are the laser electric peak amplitude, laser frequency, 
total number of cycles and carrier envelope phase 
(CEP), respectively.

The numerical algorithm used to solve the 
TDSE for our linear 1Dx1D He 
model is the Split Operator method 
described in Ref.~\cite{SplitOperator,Qfishbowl}. 
This algorithm takes advantage of the Fast Fourier Transform (FFT) paradigm to 
evaluate the kinetic energy operators of Eq.~(\ref{eq:2eham}) in 
the Fourier space. To speed up our calculations and redistribute 
the whole 2e wavefunction in position space 
--with a total number of points $N_T=N_1\times N_2\approx 4\times 10^4 \times 4\times 10^4 \approx  1.6\times10^{9}$ --
on different computational nodes, $N_p$, 
we employ the message passing interface 
MPI parallelized version of the FFTW~\cite{FFTW}. 
 This implementation allows us to reach large electron 
excursions $z_j\gg \frac{E_0}{\omega_0^2}$,
which is typical for electrons driven by spatially 
inhomogeneous 
fields~\cite{AChaconVGPRA2015,CiappinaPRA2012}.
Each 1Dx1D TDSE calculation took
about 11735 CPU-hours on $N_p=1024$ cores in 
the Barcelona Supercomputer Center.

For the He linear model, we have fixed the soft-core parameters and 
the nucleus-charge to $a=b=1$~a.u., and $Z=2$, 
respectively. With these 
values, we obtain a 2e ground state energy 
of $E_{1,2}=-2.238$~a.u. Although the matching with the experimental data is not perfect, 
it is sufficient to qualitatively reproduce the 
2e-dynamics driven by a linearly polarized 
laser-field~\cite{MLein2000,ASanpera2,ASanpera1,RGrobe1993A}. 
The 2e wavefunction~${\rm \Psi}_0(z_1,z_2)$ of the ground state  is calculated via 
imaginary-time propagation switching 
off the potential of the laser electric field in
Eq.~(\ref{eq:2elaserInhomo}), i.e. $V_{\rm int}(z_j,t)=0$. 
The inner electron and the 
outer electron ionization potentials 
are then~$I_p^{(\rm i)}=1.487$ and 
$I_p^{(\rm o)}=0.751$~a.u., respectively.

In order to follow the 2e dynamics driven by the 
plasmonic-enhanced spatially 
inhomogeneous fields, encoded in the the time-dependent
wavefunction ${\rm \Psi}(z_1,z_2,t)$, the 
wavefunction~${\rm \Psi}_0(z_1,z_2)$ of the ground state  is propagated in real time 
via TDSE with the Hamiltonian defined in 
Eq.~(\ref{eq:2eham}). In addition, we compute the 
single-electron ionization, $P_{\rm 1e}(t)$, 
and 2e-ionization, $P_{\rm 2e}(t)$, as a function 
of time $t$. The 2e position 
distribution~$|{\rm \Psi}(z_1,z_2,t)|^2$ is split into three parts: 
({\it i}) $\{|z_1|,\,|z_2| < z_a\}$, 
({\it ii}) $\{|z_1|< z_a,\,|z_2| \geq z_a\}$ or $\{|z_1| \geq z_a,\,|z_2| < z_a\}$, 
and ({\it iii}) $\{|z_1|,\,|z_2| \geq z_a\}$ with $z_a=90$~a.u.. 
The first region ({\it i}) 
describes the 2e bound wavefunction, ${\rm \Psi}_b(z_1,z_2,t)$, 
part of ${\rm \Psi}(z_1,z_2,t)$. The second one ({\it ii}) 
defines the single-electron ionization 
${\rm \Psi}_{\rm 1e}(z_1,z_2,t)$, which is the time dependent He$^+$ yield. And the third 
region ({\it iii}) includes the double-electron ionization
${\rm \Psi}_{\rm 2e}(z_1,z_2,t)$ part, which  
represents the He$^{2+}$ time dependent yield production. 
Then, by integrating over regions ({\it ii}) and ({\it iii}) the 
single- and double-electron ionization $P_{1e}(t)$ and $P_{2e}(t)$ rates (He$^{+}$ and He$^{2+}$ production yields)
are computed, respectively. 

The final two-electron momentum distribution 
${\rm S}_{\rm 2e}(p_1,p_2)=$ $|{\rm \Psi}_{\rm 2eM}(p_1,p_2,t_{\rm F})|^2$ 
is evaluated half a laser cycle after the end of the 
IR laser field as the absolute square of the 
projection of the final 2e wavepacket 
${\rm \Psi}_{\rm 2e}(z_1,z_2,t_{\rm F})$ on the 
double-electron plane waves 
${\rm \Phi}_{p_1,p_2}(z_1,z_2)=\frac{1}{2\pi}\exp{[i\left(z_1p_1+z_2p_2\right)]}$~\cite{MLein2000,CRuizPRL2008}.  
Furthermore, the correlated ion $S_{{\rm He^{2+}}}(p_{\rm ion})$ 
momentum distribution is calculated by projecting 
$S_{\rm 2e}(p_1,p_2)$ on the diagonal 
$p_1=p_2$, which corresponds to the total 2e momentum $p=p_1+p_2$. Thereby, via momentum conservation of the
system, the ion momentum reads $p_{\rm ion}= -(p_1+p_2)$.
 
In order to supplement the quantum mechanical calculations and understand the physical origin of the effects of the 
 plasmonic-enhanced field better, 
we implement CTMC simulations 
to investigate electron trajectories after
ionization of helium under the so called RIDI mechanisms. 
The simulations are restricted 
to one dimension, namely the direction of field polarization, 
in which also the field inhomogeneity develops. 
The trajectories are launched at a starting time $t_0$, which is
 distributed probabilistically following the 
 Ammosov-Delone-Krainov (ADK) 
 formula~\cite{AmmosovSP1986,DoloneJOSA1991}, 
 typically used to model strong 
 field ionization~\cite{ArissianPRL2010,LandsmanNJP2013,LandsmanJPB2014,NubbemeyerPRL2008}
\begin{align}
\begin{split}
 P(t_0,v_\perp) = \exp &\left(-\frac{2(2I_p(t_0))^{3/2}}{3 E_{\rm h}(t_0)}\right) \label{eq:ADK}
\end{split}
 \end{align}
corresponding to an atom centered 
at the origin. $I_p$ denotes the Stark shifted 
ionization potential~\cite{HofmannJPB2013}
\begin{equation}
  I_p(t_0)=I_{p}^{\rm (o)}+\frac{1}{2}(\alpha_N-\alpha_I)E_{\rm h}(t_0)^2 , \label{eq:starkShiftedIp}
\end{equation}
with $\alpha_N$ and $\alpha_I$ representing the polarizability 
of the atom and ion, respectively. The tunnel exit radius 
is assumed to be zero following the simple man's 
model~\cite{CorkumPRL1993}.
The dynamics of each electronic trajectory after ionization is solved
numerically by integrating the Newton's equations of motion, 
which takes into account the laser field, but not the Coulomb 
potential following the model in~\cite{CRuizPRL2008}.

If the electron returns to the ion ($z=0$) at time $t_{r}$ with 
kinetic energy $E_{k1}(t_{r})$ larger than the ionization potential
$I_{p}^{\rm(i)}$ of the second electron~\cite{ASanpera1}, this second, inner electron can be ionized as well. In this ionization process
the kinetic energy of the first electron is reduced by $I_{p}^{(\rm i)}$ 
and the second electron is born in the continuum with zero velocity 
\begin{align}
\begin{split}
 p_1(t_{r}) &= \pm \sqrt{2\left(E_{k1}(t_{r}) - I_{p}^{(\rm i)}\right)}\\
 p_2 (t_{r}) &= 0  \label{eq:p1Andp2AfterDoubleIonization}.
\end{split}
\end{align}
Here, the two different signs in $p_1$ describe the possibility 
of scattering the first electron into forward, binary, 
or backward, recoil, direction with
respect to its momentum directly before the
ionization of the inner electron. 
For each double ionization event,
we calculated both options. The dynamics after the second
ionization is again determined by the propagation in the laser field, 
where the Coulomb force is completely 
neglected~\cite{CRuizPRL2008}.

\section{Double-electron ionization }\label{sec:2}
To study the e-e correlation effects we firstly compute the final 
single- and double-electron ionization yields as a function of the 
peak laser field intensity for a few-cycle IR pulse. This 
allows us to identify the intensity regions 
where the spatially inhomogeneous field substantially 
modifies the double-electron ionization process. Secondly, we compute the 
two-electron momentum distribution as 
a function of the inhomogeneity degree at a fixed intensity. This scan on $\beta$ 
provides enough evidence about the 
role of the inhomogeneous field 
in the 2e ionization process. Furthermore, we scrutinize if any e-e correlation effects
can be found in this process.   

\subsection{He$^+$ and He$^{2+}$ ion yields}
\begin{figure*}
\center
\resizebox{0.5\textwidth}{!}{ \includegraphics{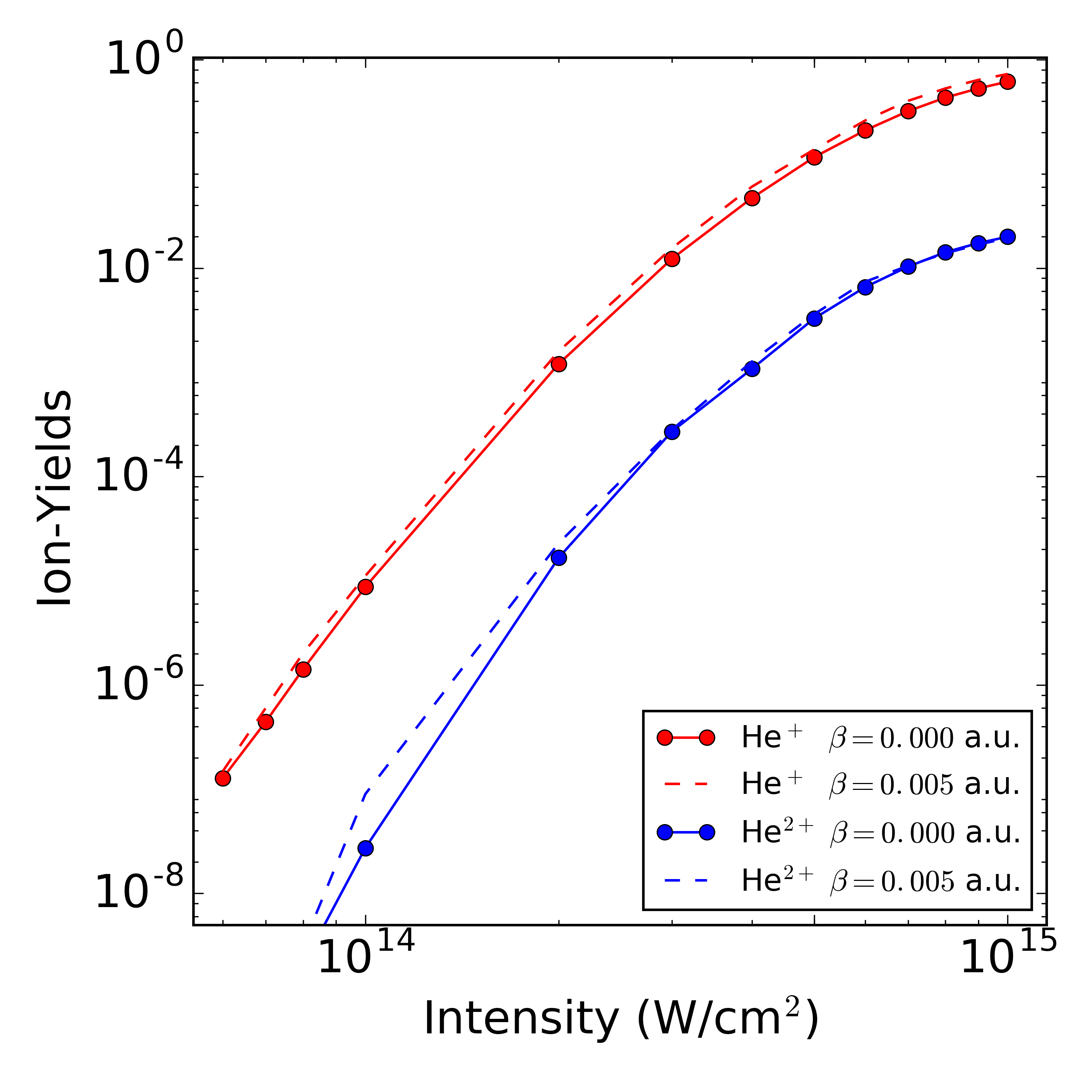} }
\caption{Numerical TDSE calculations of the single- (red) and 
double-electron (blue) ionization yields of our linear 
1Dx1D He 2e-model driven by homogeneous (line with circles) 
and inhomogeneous (dashed line) fields as a function of the 
laser peak intensity. The mean frequency of the IR laser 
field is $\omega_0=0.057$~a.u. (1.55~eV), 
the CEP is $\varphi_{\rm CEP}=0^{\circ}$ and the total number 
of cycles is ${\rm N}=4$ under a $\sin^2$ envelope.}
\label{fig:1}       
\end{figure*}

We numerically compute the final 2e-ionization yield by  
the procedure described in Section~2. The grid parameters 
used in those calculations are 
$N_1=\,N_2=40960$ points 
and $\delta z_1=\delta z_2=0.25$~a.u. The integration time step 
was chosen $\delta t=0.025$~a.u. The results of the single- 
and double-electron ionization 
yield as a function of the peak laser field intensity for 
the homogeneous ($\beta=0$) and inhomogeneous fields with  
$\beta=0.005$~a.u., are depicted in Fig.~\ref{fig:1}.  

A sizeable enhancement of the final 2e 
ionization $P_{\rm 2e}(t_{\rm F},I_0)$ is observed 
for the inhomogeneous field case 
when compared to the conventional one. Similar effects are also
obtained in the comparison with the single-electron ionization yields. 
This enhancement clearly shows that the spatially 
inhomogeneous fields play an instrumental role in the NSDI
of helium. 

\begin{figure*}
\resizebox{0.5\textwidth}{!}{\includegraphics{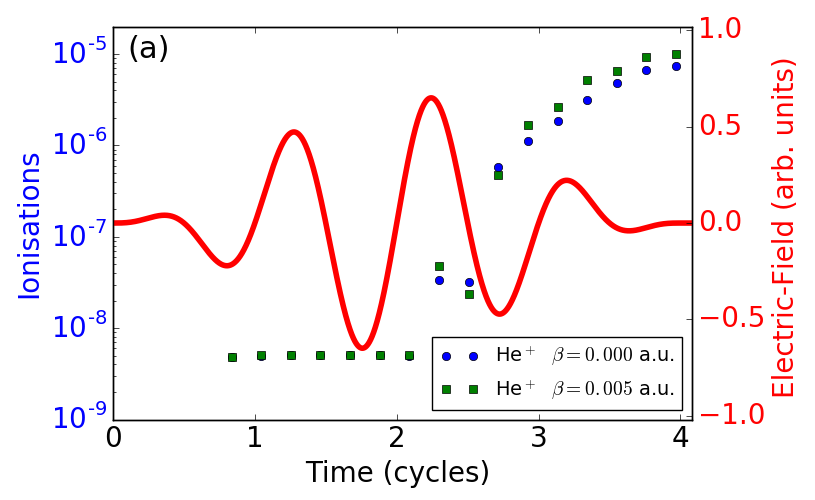} }
\resizebox{0.5\textwidth}{!}{\includegraphics{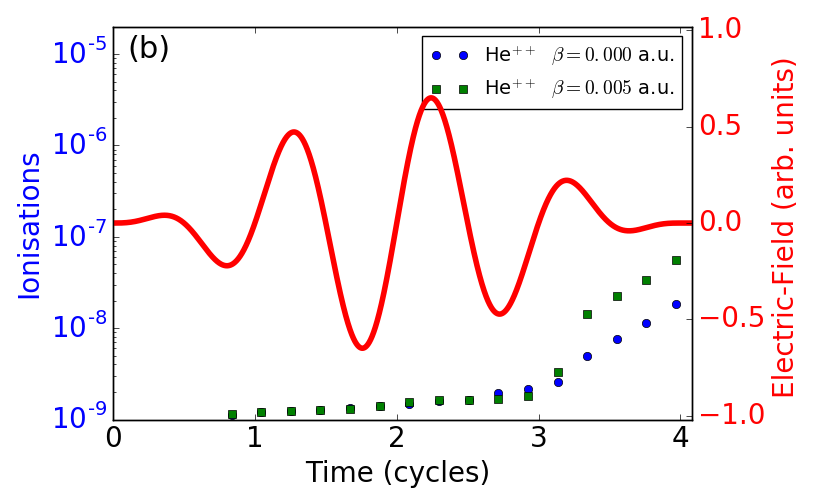} }
\caption{Single- (a) and double-electron (b) ionization 
yields of our linear 1Dx1D He 2e-model driven by conventional
and spatially inhomogeneous fields as a function time 
(see left axis). The IR laser field oscillations are depicted in 
red solid line. The peak intensity used to follow the two-electron
dynamics is $I_0=2\times10^{12}$~W/cm$^2$. The other laser 
parameters are the same than those used in Fig.~\ref{fig:1}. }
\label{fig:2}  
\end{figure*}

Naturally, the question about the origin of this enhancement arises. 
In order to answer it, we compute the 
single- $P_{\rm1e}(t)$ and double-electron
$P_{\rm e2}(t)$ ionization yield as a function 
of time at a fixed peak intensity of $I_0=2\times10^{14}$~W/cm$^2$.
Here we focus our attention on the intensity 
region where the double ion yield, He$^{2+}$, is enhanced by 
the inhomogeneous field. According 
to Fig.~\ref{fig:1}, one such region is 
$I_0$$=\,1-5\times10^{14}$~W/cm$^2$. 
The results of the time-evolved probabilities 
are depicted in Fig.~\ref{fig:2}. 

For the single-electron ionization $P_{\rm 1e}(t)$ shown in 
Fig.~\ref{fig:2}(a), the ``inhomogeneous" ionization 
yield is larger than the conventional one, in particular, at about 2.5 
cycles of the IR laser. 
We could trace out the origin of this observation in a 
much stronger distortion of the laser-atomic potential barrier, which raises
the probability of the first bound electron to 'escape' from the atom.

Fig.~\ref{fig:2}(b) shows a comparison of $P_{\rm 2e}(t)$ 
for conventional and inhomogeneous fields. About 3.4
cycles of the IR laser oscillations, the 2e 
ionization yield largely increases for the inhomogeneous field
case with respect to the conventional one by more than 5-times. 
At this very low inhomogeneity degree of $\beta=0.005$~a.u., 
and low IR peak intensity, this enhancement of the 
 2e-ionization rate is a very surprising result.  
Similar behaviour was previously observed 
in~\cite{AlexisHHG2e}, where the double-electron
ionization reaches higher yields leading to an enhancement 
in the intensity of the HHG signal. However, in that latter 
case a larger inhomogeneity degree of $\beta=0.02$~a.u was used. 

An hypothesis that might explain that result is based on the 
three step Corkum's model~\cite{CorkumPRL1993,WBeckerRMP2012,BWalkerPRL1994} 
where (1) the first electron ionizes via tunnelling, (2) then this electron 
propagates in the continuum gaining energy from the laser field 
 - in our case a spatially inhomogeneous field - 
 and then when the field changes its sign the electron 
 has a probability to re-collide with the ion core He$^+$. 
As a third step, this colliding electron can kick out the second 
inner electron if and only if the first electron kinetic energy is larger
than $I_p^{\rm(i)}$, the ionization potential of the second 
inner electron. This process is called double-ionization by 
re-scattering impact direct ionization (RIDI) or (e, 2e). 

From the behaviour of electrons driven by spatially inhomogeneous fields (see e.g.~\cite{reviewROP}), it is very likely that the first-ionized electron gains a much larger energy compared to the conventional case. Thus, 
at the instant of recollision, the second electron would have a higher chance to be ionized in a spatially inhomogeneous field, which corresponds to an enhancement of the double electron ionization probability. 

\begin{figure*}
\resizebox{0.5\textwidth}{!}{\includegraphics{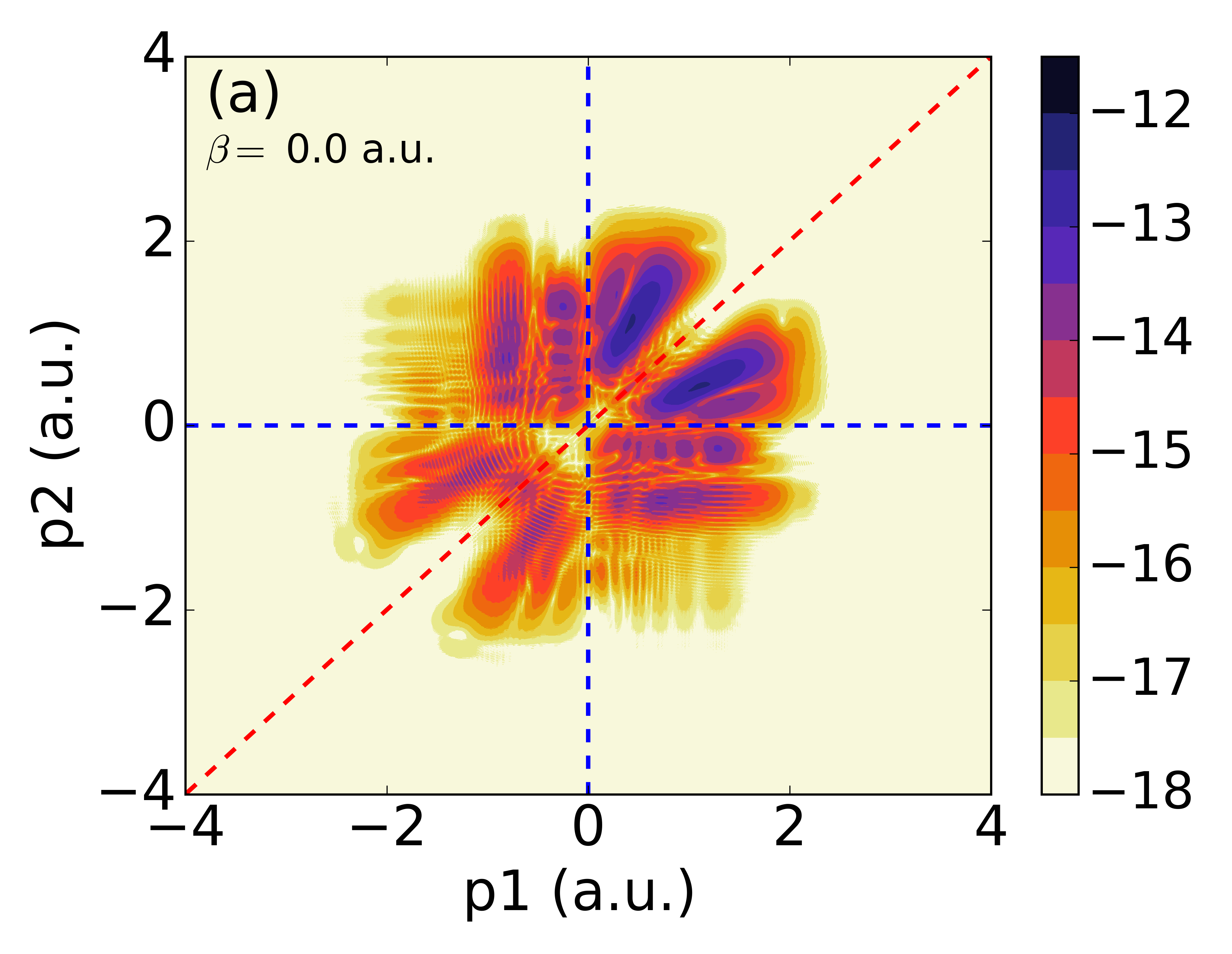}}
\resizebox{0.5\textwidth}{!}{\includegraphics{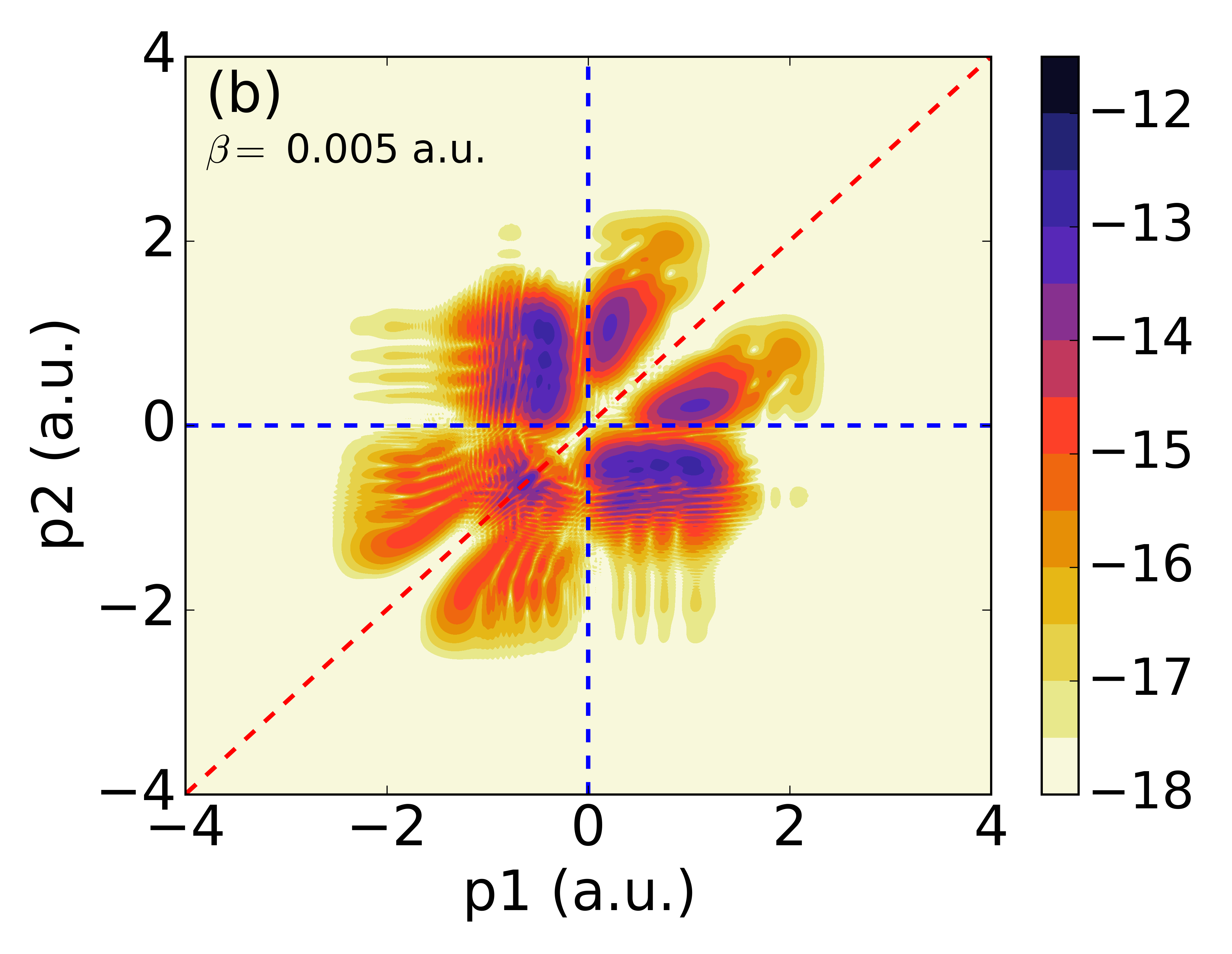} }
\resizebox{0.5\textwidth}{!}{\includegraphics{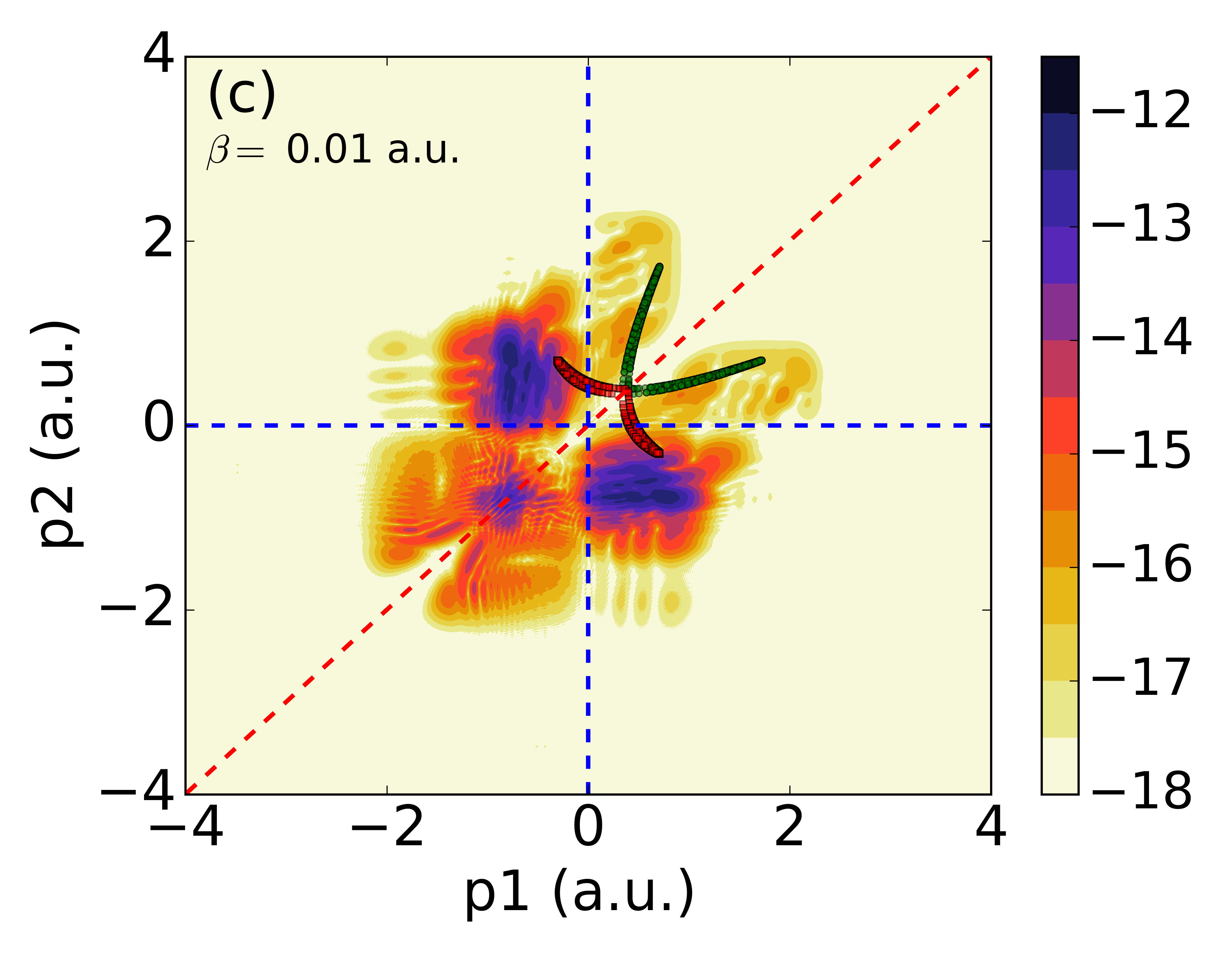} } 
\resizebox{0.5\textwidth}{!}{\includegraphics{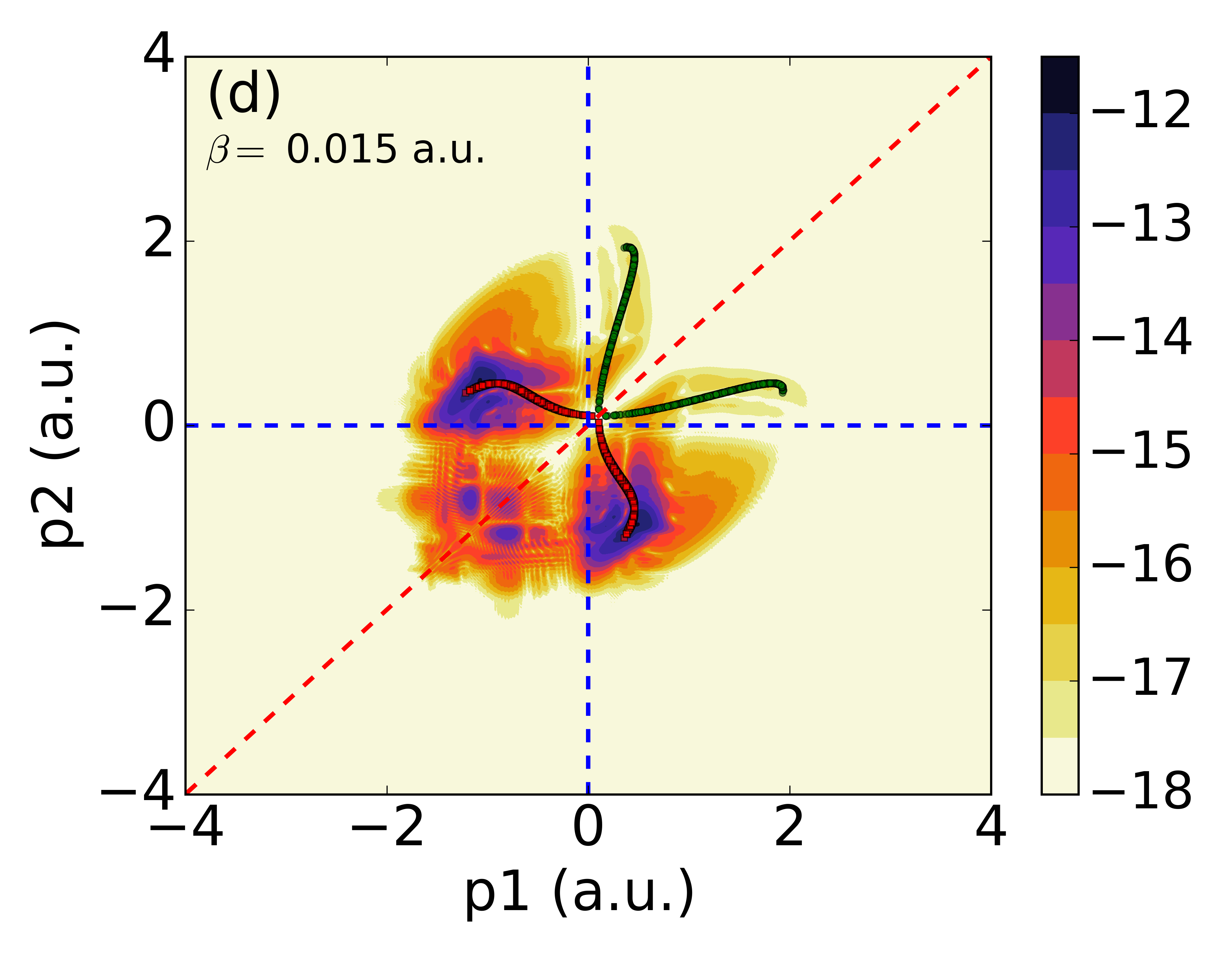} }
\caption{Numerical two-electron momentum distribution for various 
inhomogeneity degrees: $\beta=0.000$~(a),
$0.005$~(b), $0.010$~(c) and $0.015$~a.u.~(d). The color
scale is $\log_{10}[S_{\rm 2e}(p_1,p_2)]$. Vertical and horizontal
blue dashed lines denote the 2e momentum axes which help us 
to distinguish between correlated, (i) and (iii) quadrants, and 
anticorrelated, (ii) and (iv) quadrants, regions. The diagonal red 
dashed line $p_1=p_2$ represents the max e-e correlation 
momentum points or the total 2e momentum $p=p_1+p_2$. CTMC 
for the (e, 2e) mechanisms are superimposed in green-circles
(recoil process) and in red-squares (binary process) 
in (c) and (d) panels. 
The laser-peak intensity used for these numerical calculations is $I_0=2\times10^{12}$~W/cm$^2$. 
The other laser parameters are the same as 
those used in Fig.~\ref{fig:1}.}
\label{fig:3}     
\end{figure*}

\subsection{Correlated two-electron momentum maps }
Another interesting observable, which contains information 
about the e-e correlation, is the 2e-momentum distribution. This observable
has allowed to disentangle the common sequential and non-sequential
double RESI, rescattering impact ionization and laser-field assisted 
rescattering ionization mechanisms~\cite{KopoldPRL2000,EreminaJPB2003,LiuYPRL2008}. 
Fig.~\ref{fig:3} depicts $S_{\rm 2e}(p_1,p_2)$ 
for different $\beta$ parameters at the same 
fixed laser peak intensity of $I_0=2\times10^{14}$~W/cm$^2$. 
The double-electron map in Fig.~\ref{fig:3}(a) exhibits two large probability
peaks on the first quadrant of the correlation region - in almost perfectly concordance with 
the data published in ref.~\cite{MLein2000}. 
This probability distribution indicates that 
both electrons prefer to leave on the same (positive) direction. 
It is understood that the repulsive e-e Coulomb potential plays 
an important role at those relative low peak intensity for 
the He model~\cite{MLein2000}.

Note that a classical rescattering electron 
scenario (e, 2e) is not good enough for describing this NSDI mechanism  
of our He model at this peak intensity. From a classical viewpoint, 
the rescattering energy $E_{\rm k,max}=3.17U_p=1.4$~a.u.,
is lower than the ionization potential of the 
inner electron $I_p^{\rm(i)}\sim1.5$~a.u. 
Instead, this double-electron ionization map
 could be understood as a laser-field-assisted
rescattering process for which such a constraint does not 
apply~\cite{CRuizPRL2008,KopoldPRL2000,EreminaJPB2003,MLein2000,LiuYPRL2008}.  
As pointed out in~\cite{WBeckerRMP2012,MLein2000}, 
the driving laser field provides the rest of the required energy to 
remove the second electron at the instant of recollision.

For further interpretations of Fig.~\ref{fig:3} we recall that 
finding double ionization in 
quadrants I and III corresponds to both electron momenta pointing in the same direction. In contrast, quadrants II and IV 
contain the cases of the electrons' momenta pointing into opposite directions. 
When both electrons leave the atom in the same direction, we say they are correlated.
Comparing the 2e-momentum distributions in Figs.~\ref{fig:3}(a) and (b), we find that the two 
electrons prefer to detach in opposite directions when driven by plasmonic-enhanced spatially inhomogeneous fields.  
This effect is even larger for an inhomogeneity degree of 
$\beta=0.01$ and $0.015$~a.u., as can be seen 
in Figs.~\ref{fig:3}(c)-(d). We note, however, the appearance of a small 2e-probability also in the correlated regions.

Naturally, questions about the physical 
mechanisms behind those effects
in the 2e maps emerge.
In order to address those questions, we superimposed our
CTMC calculations on the TDSE results in Figs.~\ref{fig:3}(c) and (d) for the cases of
binary (red-squares) and recoil (green-circles) processes. 
For those results, an excellent agreement between the 
TDSE and the CTMC calculations is found, in particular for the case of an inhomogeneity degree 
$\beta=0.015$~a.u.. 
This clearly corroborates that
the forward rescattering process with respect to the first 
incident electron direction, binary, is highly probable within 
that so-called (e, 2e) mechanisms if 
spatially inhomogeneous fields drive the two-electron system. 
Note that this agreement of TDSE and 
CTMC supports our previous observation that the 2e-particles are
likely to prefer to leave the atom in opposed directions. \\
According to Weber et al.~\cite{WeberNat2000} the momentum 
distribution corresponding to the coordinates 
$p=p_1+p_2$ (diagonal along $p_1=p_2$) and
 $p^-=p_1-p_2$ (diagonal $p_1=-p_2$), are helpful for describing
 the importance of two effects: e-e repulsion and acceleration of 
 the particles by the optical field. On the one hand, e-e repulsion
 does not change $p$ but contributes to $p^-$. On the other
 hand, the momentum transfer received from the field is identical. 
So, this part of acceleration does not change $p^-$ but contributes 
to $p$. Note, however, that this statement is only valid if the 
electric field does not depend on the position. Thereby, for the 
inhomogeneous field cases we cannot conclude that the acceleration 
part does not contribute to $p^-$. This is in absolute
concordance with what we observe in the 2e momentum maps 
for $\beta=0.010$ and $0.015$~a.u..

\begin{figure*}
\resizebox{0.5\textwidth}{!}{\includegraphics{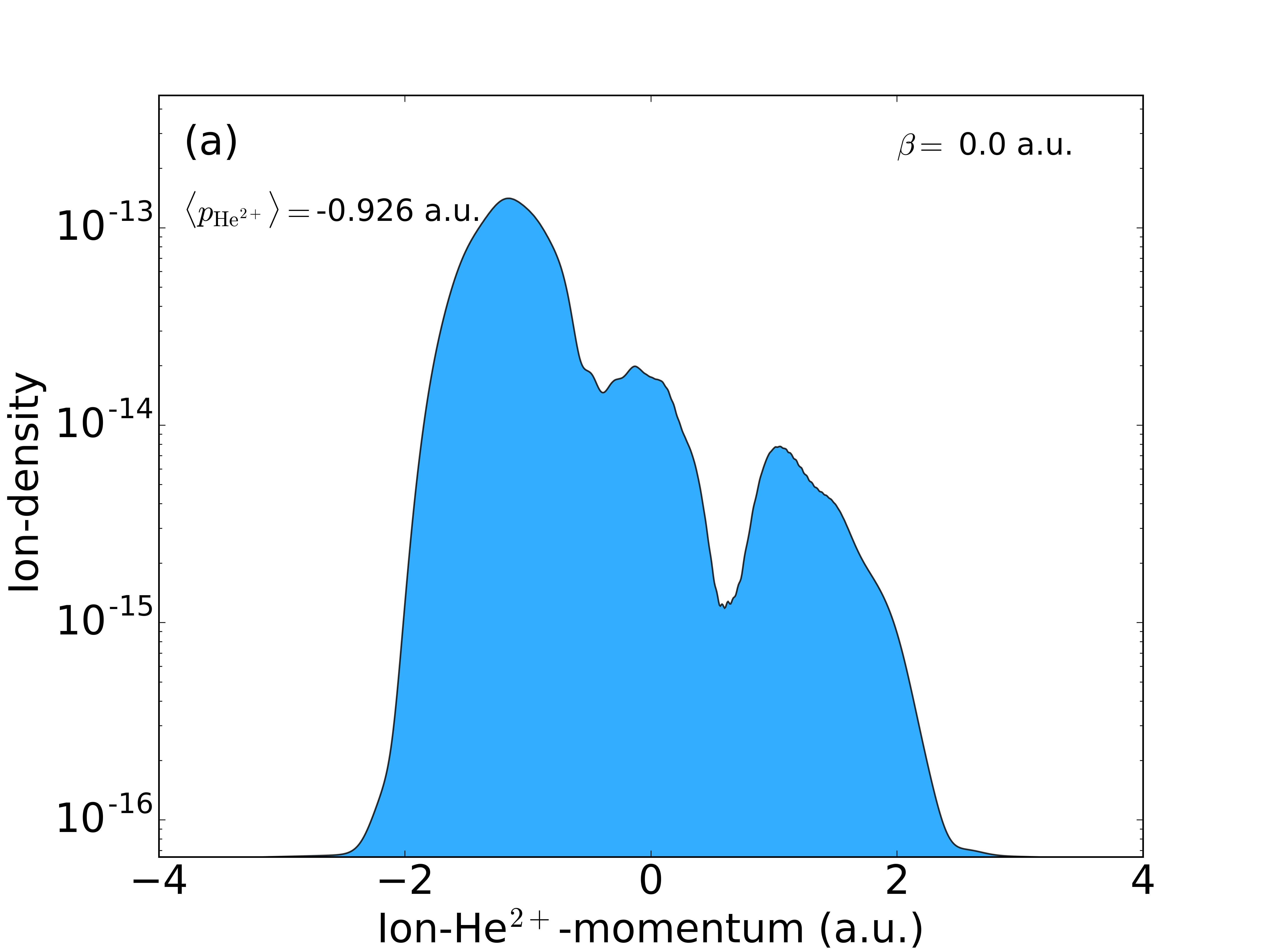} }
\resizebox{0.5\textwidth}{!}{\includegraphics{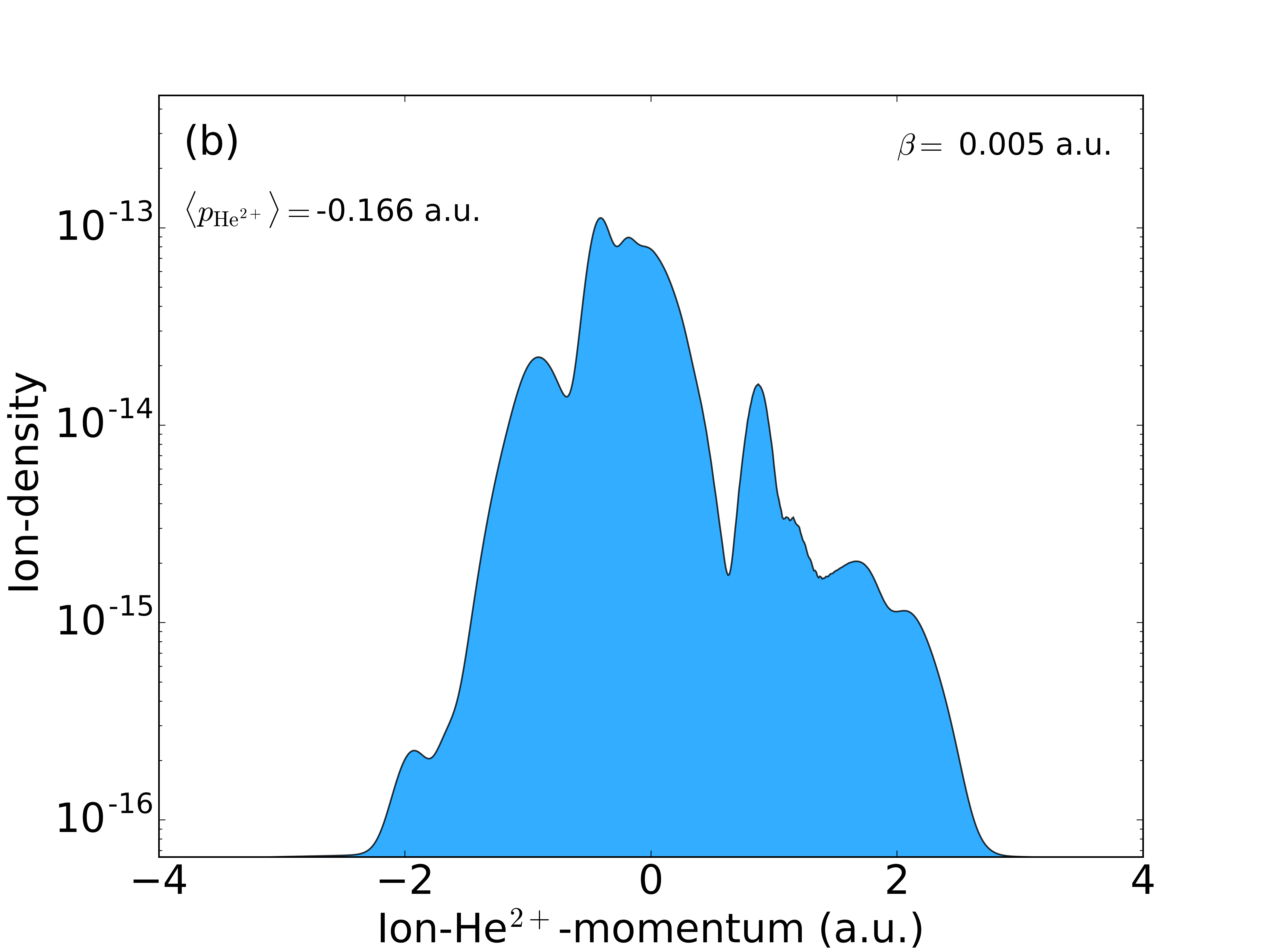} }
\resizebox{0.5\textwidth}{!}{\includegraphics{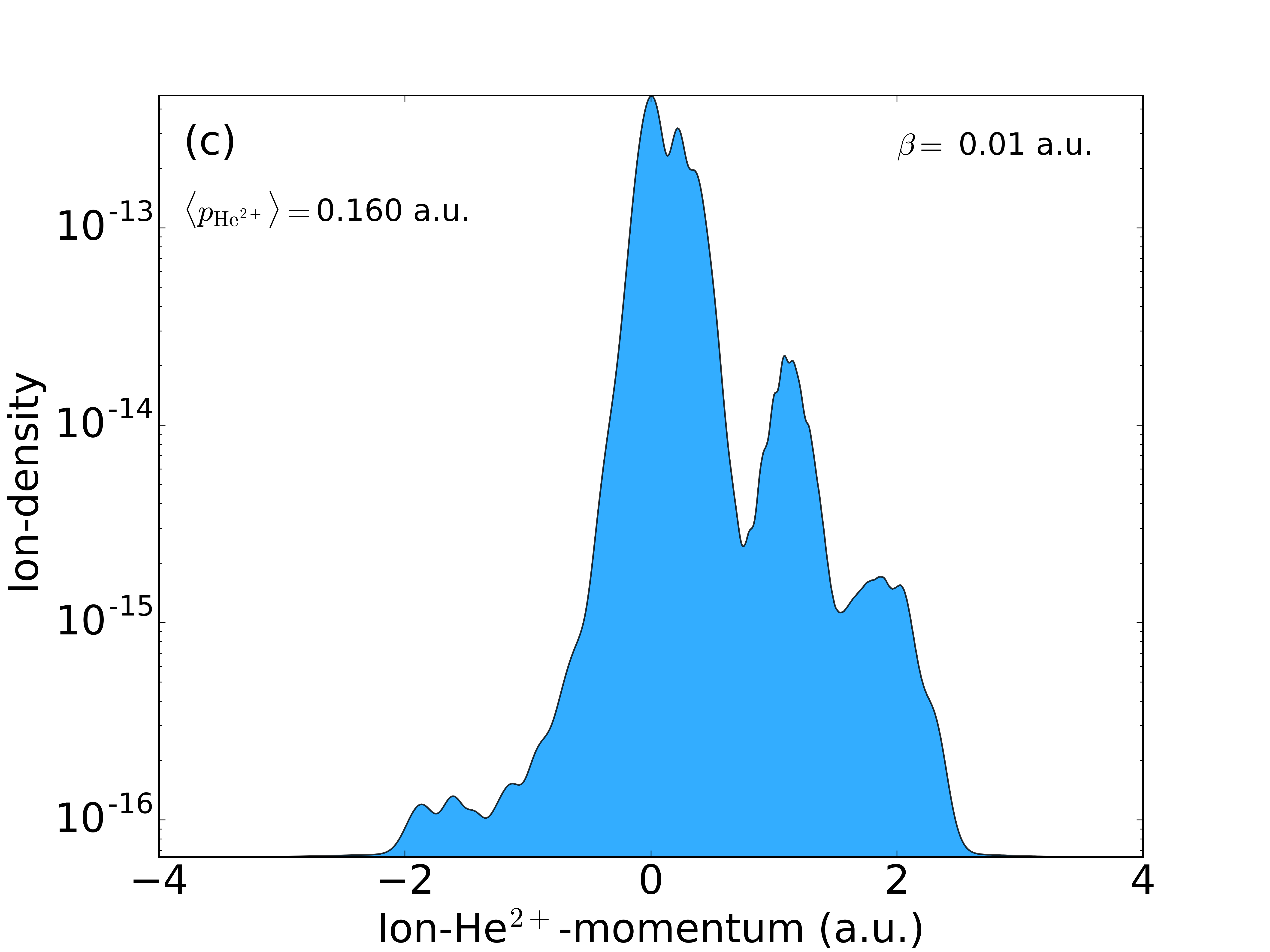} }
\resizebox{0.5\textwidth}{!}{\includegraphics{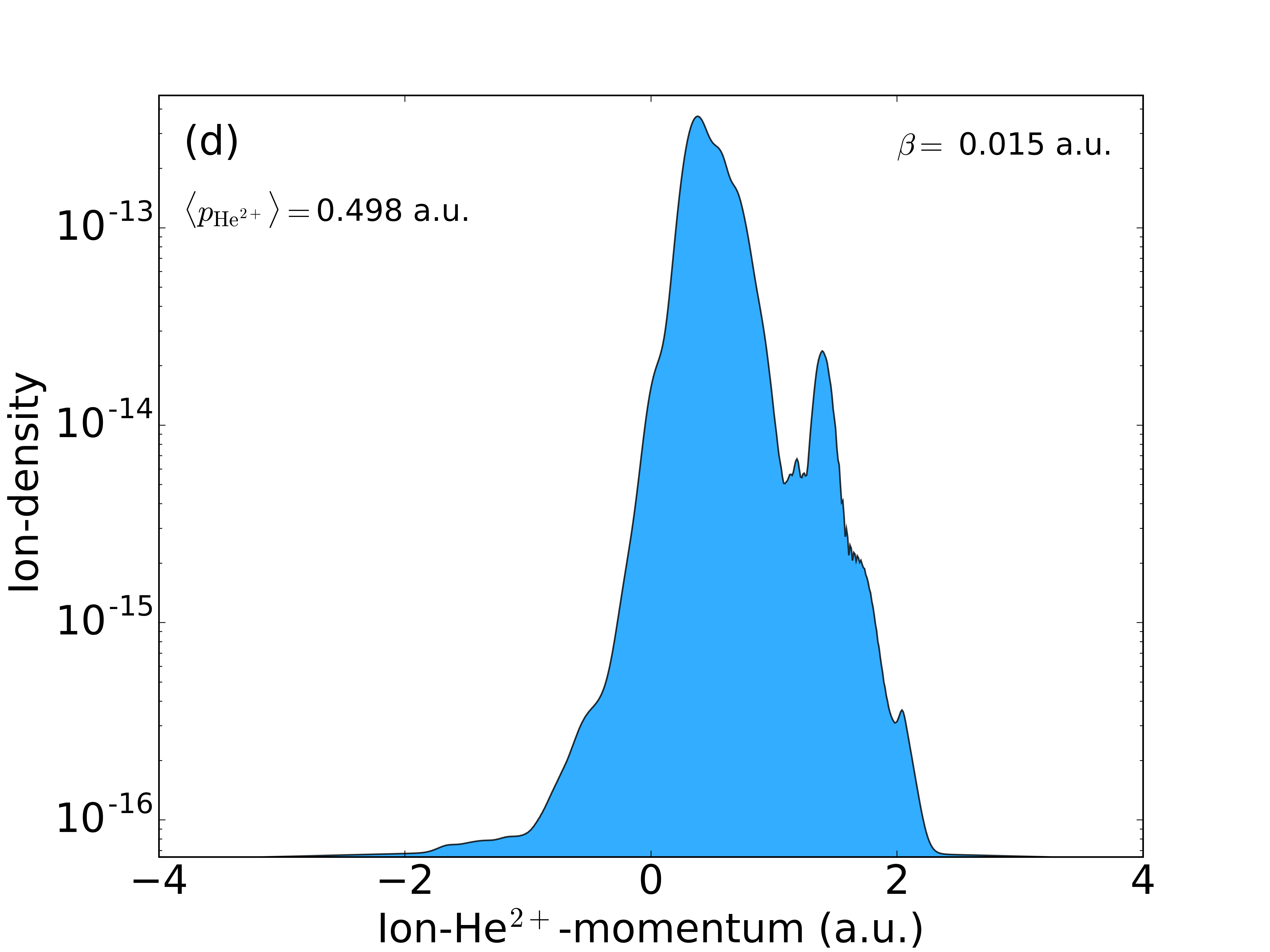} }
\caption{Correlated ion He$^{2+}$ momentum 
distributions, $S_{{\rm He^{2+}}}(p_{\rm ion})$, corresponding to the panels (a)-(d) of Fig.~\ref{fig:3}.}
\label{fig:4} 
\end{figure*}

Additionally, in Fig.~\ref{fig:4} we show the 
correlated ion He$^{2+}$ momentum distributions corresponding to the 
$S_{2e}(p_1,p_2)$ panels of Fig.~\ref{fig:3}. 
A first observation is that a sizeable 
momentum-shift is found for the 
recoiling ion as the inhomogeneity 
degree $\beta$ increases. 
For the conventional field case depicted in Fig.~\ref{fig:4}(a), 
the full momentum width of the distribution is about 
$\pm2A_0=\pm2.6$~a.u., where $A_0=E_0/\omega_0$ is 
the maximum vector potential for the conventional 
field~\cite{WeberPRL2000,MLein2000,RudenkoPRL2004}. 
An asymmetry in the amplitude of the ion distribution 
$S_{{\rm He^{2+}}}(p_{\rm ion})$ is observed at 
$\beta=0$~a.u. This is due to the employed 
laser-field being within the few-cycle regime, ${\rm N}= 4$, 
see e.g. Ref.~\cite{BerguesIEEE2015} about the CEP 
effects. 
However, three peaks at about
$p_{\rm ion}^{\rm (max)}=\{-A_0,\,0,\, +A_0\}$ 
are found. These might suggest that the laser-field assisted 
rescattering double ionization mechanism and the 
RESI mechanism take 
place simultaneously in 
such a complex correlated 
momentum map.

In case of inhomogeneous fields, the ion
distribution shape strongly depends on 
the parameter $\beta$. While the inhomogeneity 
increases, the expectation value of the ion momentum 
 $\langle p_{\rm He^{2+}} \rangle$ is shifted from negative
to positive momentum, indicating that the ion recoils in the completely
opposite direction compared to the conventional field case. 
Note that these several peaks that appear in the ion-distributions, 
suggest the possibility of different interference 
paths in the processes of DI driven by that spatially 
inhomogeneous field.  
 
\begin{figure*}
\resizebox{0.45\textwidth}{!}{\includegraphics{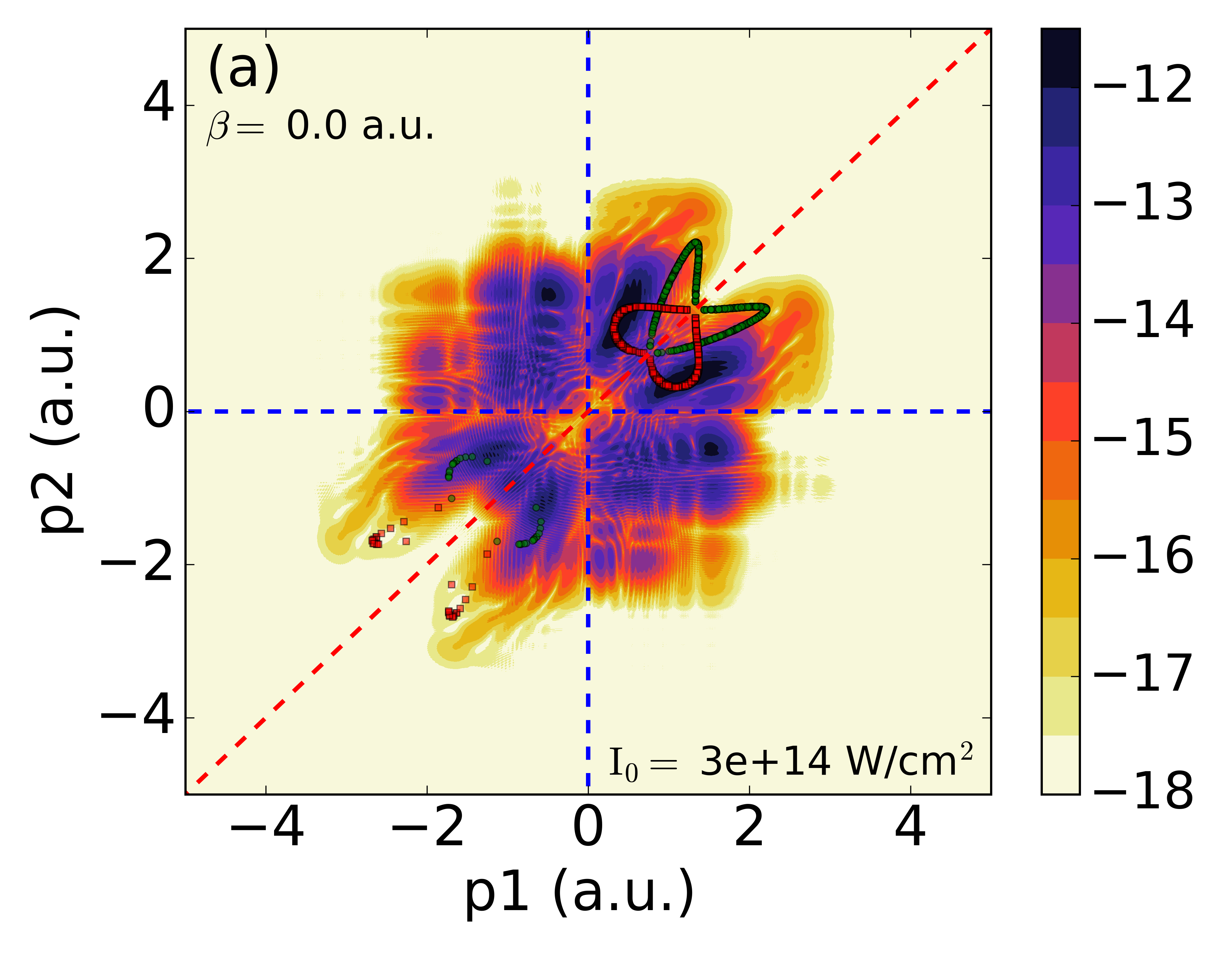}}
\resizebox{0.45\textwidth}{!}{\includegraphics{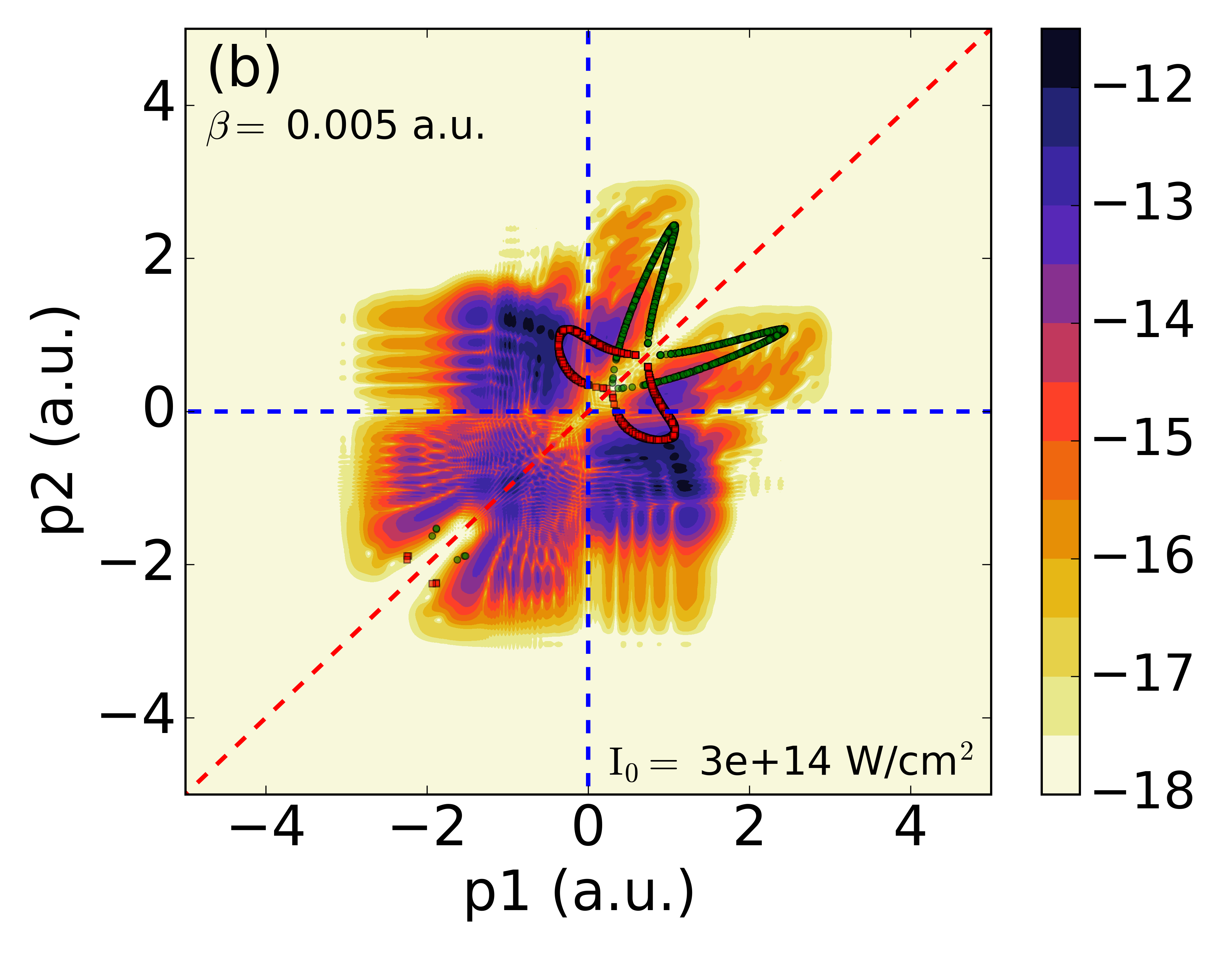}}
\resizebox{0.45\textwidth}{!}{\includegraphics{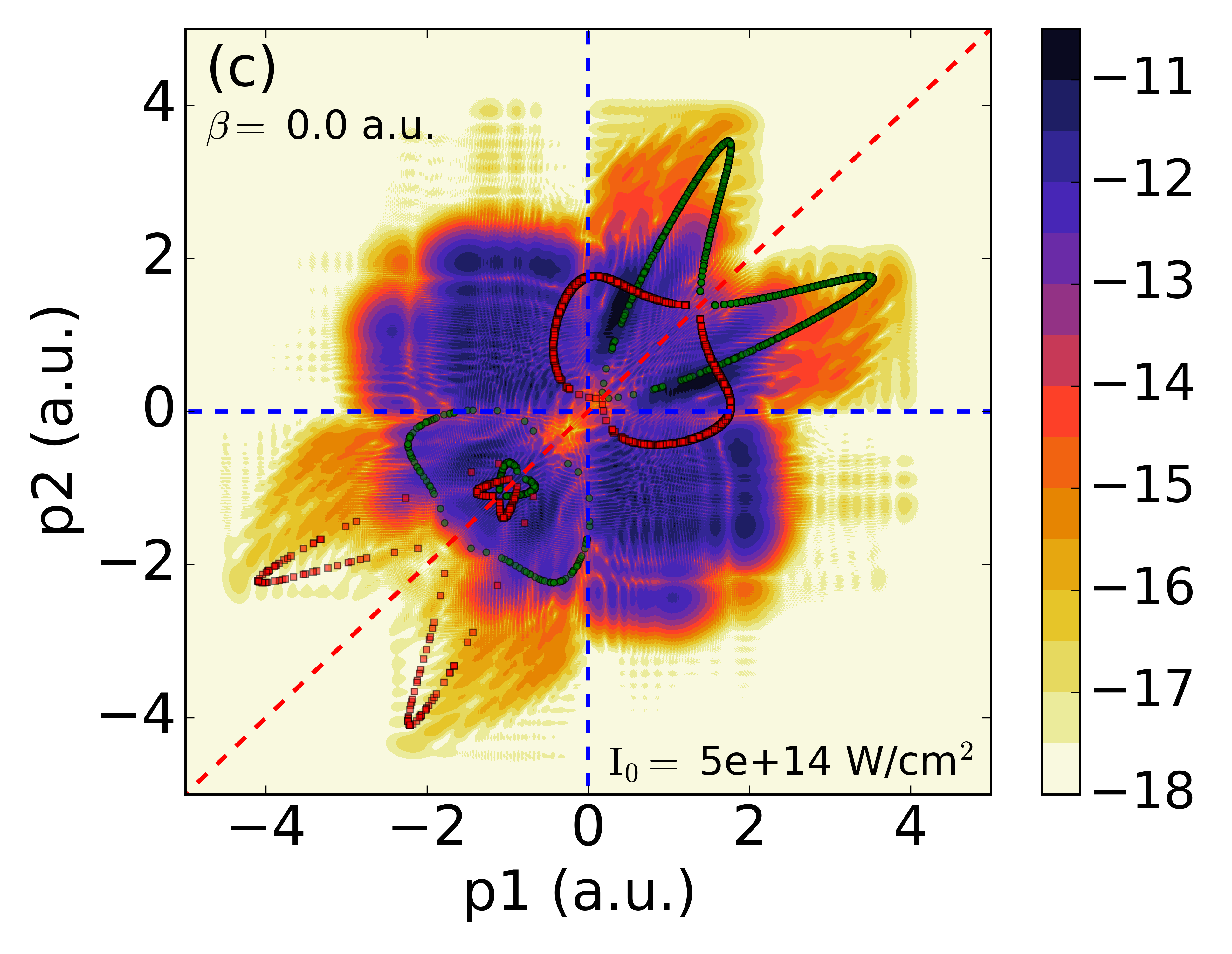}}
\resizebox{0.45\textwidth}{!}{\includegraphics{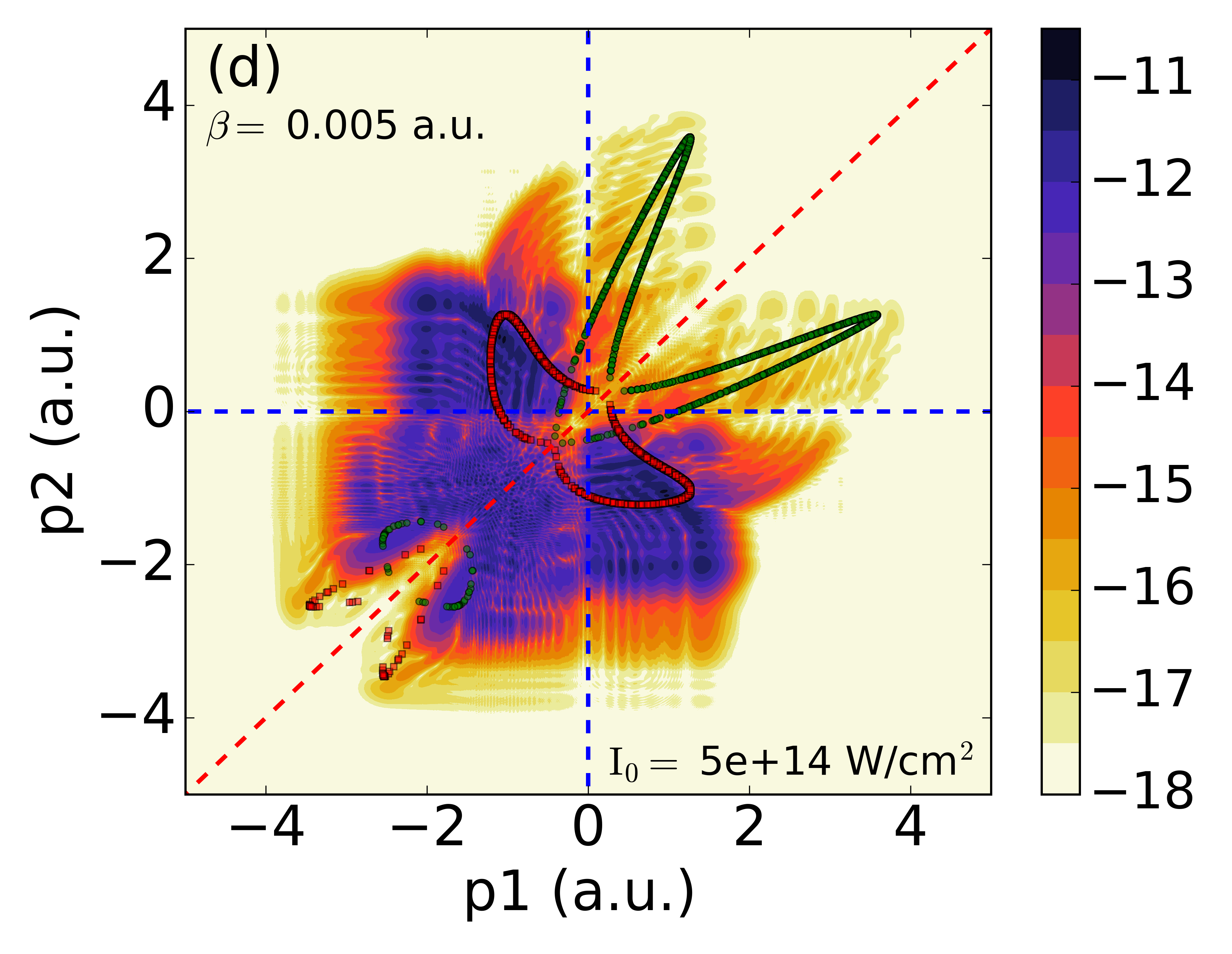}}
\resizebox{0.45\textwidth}{!}{\includegraphics{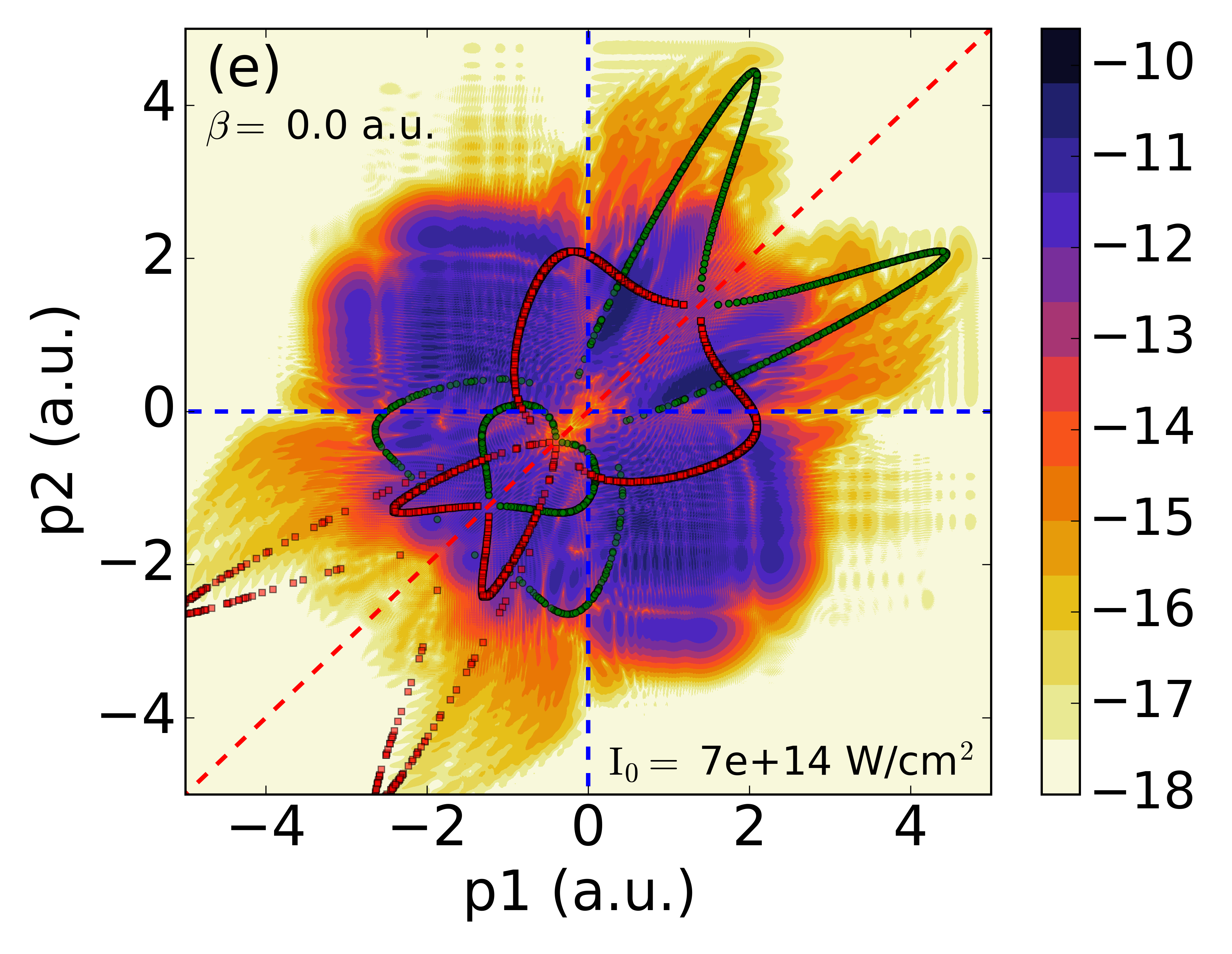}}\hspace{1cm}
\resizebox{0.45\textwidth}{!}{\includegraphics{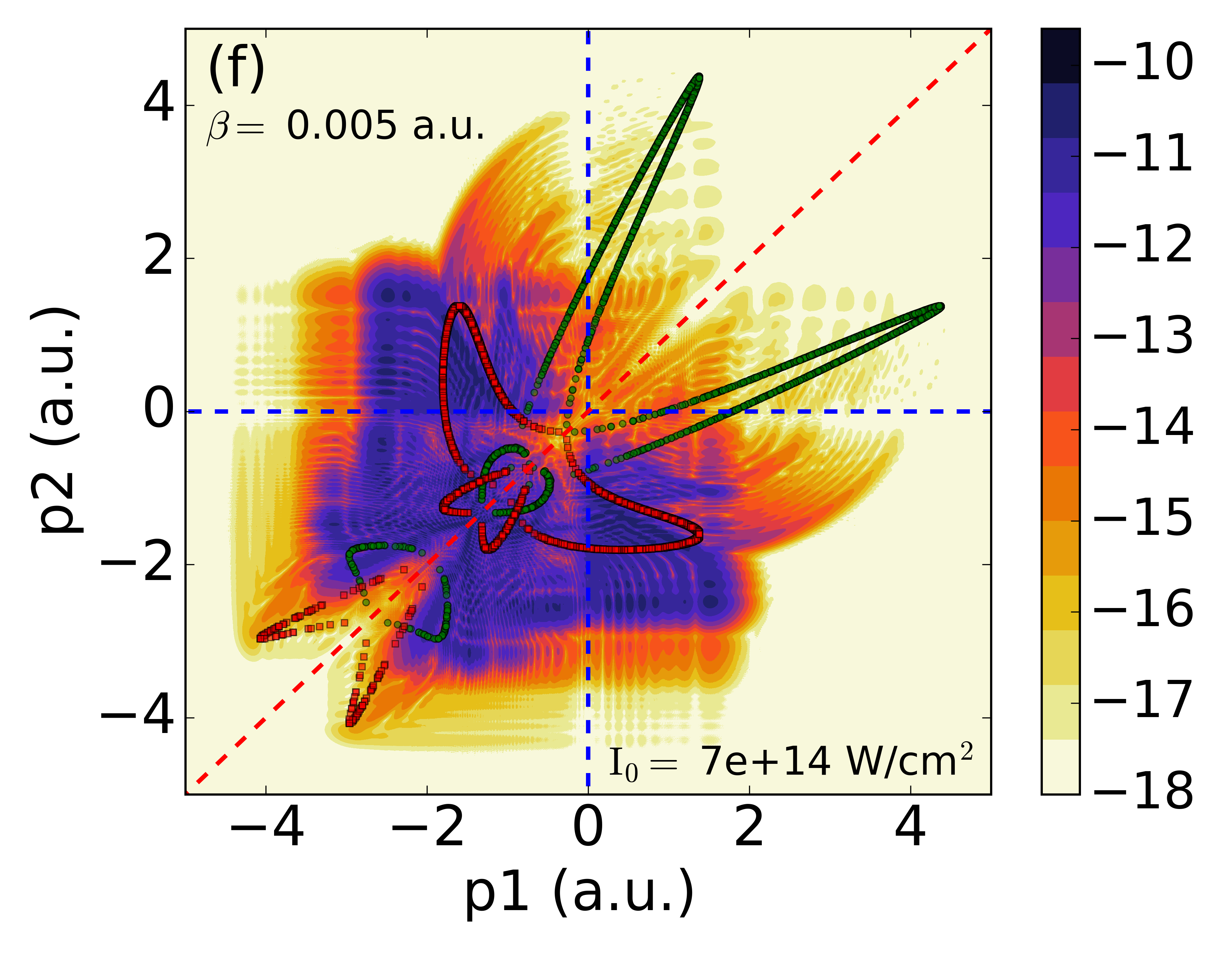}}
\caption{Numerical two-electron momentum distributions driven by
 homogeneous $\beta=0$ (a), (c), (e), and inhomogeneous fields (b), 
(d), (e) with $\beta=0.005$~a.u., for three different 
laser-peak intensities: $I_0=\, 3, \, 5, \, 7\times10^{14}$~W/cm$^2$. 
The CTMC calculations for binary (red-squares) 
and recoil (green-circles) processes are 
superimposed on the momentum-momentum maps. 
Other laser parameters are the same as in Fig.~\ref{fig:1}. }
\label{fig:5} 
\end{figure*}

So far, we have studied double ionization in He via
scanning the 2e-momentum distribution over the inhomogeneity 
parameter $\beta$ at a fixed laser intensity. In order to obtain an insight 
about the 2e ionization when the laser-peak 
intensity increases, we compute and compare the 
momentum-momentum distributions 
for the conventional $\beta=0$ and inhomogeneous 
$\beta=0.005$~a.u fields. Additionally, our {\em ab}-initio 
TDSE calculations are compared with the CTMC simulations. 
The results are depicted in Fig.~\ref{fig:5}. \\
While the peak intensity increases from 
$3$ to $7\times10^{14}$~W/cm$^2$ for conventional fields, 
some pronounced lobes in the correlation regions are observed. 
Those density lobes are larger in the anticorrelated region.
This is a signal that the e-e Coulomb repulsion force 
is losing its importance while the 
laser-field peak intensity increases. In particular, 
that effect is larger for the highest intensity. 
In addition, note that a better agreement between TDSE and 
CTMC is found in the cases of Figs.~\ref{fig:5}(c) and (e) 
as it is expected~\cite{MLein2000,WeberPRL2000,CRuizPRL2008}. 
This indicates that the (e, 2e) processes are the main mechanisms 
behind those calculations.  
However, in Fig.~\ref{fig:5}(a), a laser-field assisted rescattering
DI process still dominates over the RIDI mechanisms. 
This we conclude from the poor agreement between
the quantum mechanical and the classical calculations 
for the binary and recoil processes. 

On the other hand, for inhomogeneous fields, Fig.~\ref{fig:5}(b),
(d) and (f), the probability of 2e ionization with opposite momenta 
increases. This clearly indicates that the propagation of electrons 
under the influence of plasmonic field changes completely the 
2e-dynamics. Note, that a signal in quadrant III of the
correlation region is also observed, which is an indication
that both electrons, independently of the incident direction 
of the first colliding particle, prefer to leave with negative 
momenta directions. 
Furthermore, while the laser peak intensity increases, 
the V-like shape in the quadrant III tends to be much closed, 
and also a strong  signal along the diagonal $p_1=p_2$ for $p_1<0$
is clearly observed. These facts are the signature of e-e correlation
effects rapidly losing importance while the particles are propagating in
the plasmonic field. Note, however, that e-e repulsion somehow is still
present because of the large momentum density 
width along the diagonal $p_1=-p_2$. 

Finally, it is interesting to point out that up to 
$7\times10^{14}$~W/cm$^2$
the NSDI by inhomogeneous fields is still 
within the rescattering (e, 2e) scenario. This statement is
supported by the CTMC simulations that agree very well with the TDSE calculations for all studied cases. 
This demonstrates that the isolation of binary and recoil processes
is very sensitive to the laser peak intensity. We should note, however, that between 2 
and 5$\times10^{14}$~W/cm$^2$, we ensure that those backward
and forward rescattering processes could be separated, just by 
observing the anticorrelating and correlating regions of the
momentum-momentum distribution. 

\section{Conclusions}
Non-sequential double ionization of helium atoms driven 
by a plasmonic-enhanced spatially
inhomogeneous fields has been theoretically investigated. 
By means of the fully  numerical solution of the 
time-dependent Schr\"odinger equation,
we observed that ion yield of He$^{2+}$ substantially 
increases while the 
inhomogeneous field drives the system. 
 An analysis of the single- and double-electron 
time-evolution probabilities and the two-electron 
momentum distribution simulations of the binary an 
recoil mechanisms support 
that the main reason for this 
enhancement corresponds to a high accumulated energy of the 
first re-colliding electron when it is moving in the spatially
inhomogeneous field. 
 
An unexpected (e, 2e) mechanism at very low intensity, 
i.e. $I_0=2\times10^{14}$~W/cm$^2$ 
is observed with increasing inhomogeneity. 
This means that by engineering the 
inhomogeneous field the two different mechanisms, namely the 
laser-field-assisted re-scattering and the RIDI process can be isolated. 
Furthermore, our interpretation of the fully ab-initio TDSE for the 
two-electron momentum distributions 
by comparing to CTMC simulations, allowed us 
to distinguish between binary and recoil processes if and only if 
the spatially inhomogeneous field drives the system.

Still there are open questions, e.g. concerning the role of the e-e 
Coulomb potential while both identical particles are propagating 
within the spatially inhomogeneous field and how this effect is related
to the 2e momentum distribution maps for larger laser peak 
intensities here used. 
We plan to address these questions in subsequent work.

\section{Acknowledgements}
This work was supported by the project ELI--
Extreme Light Infrastructure--phase 2 (Project No. CZ.02.1.01/0.0/0.0/15\_008/0000162) from European
Regional Development Fund, Ministerio de Econom\'ia
y Competitividad through Plan Nacional [Grant No.
FIS2011-30465-C02-01, FrOntiers of QUantum Sciences
(FOQUS): Atoms, Molecules, Photons and Quantum
Information Grants No. FIS2013-46768-P and No.
FIS2014-56774-R; and Severo Ochoa Excellence Grant
No. SEV-2015-0522], the Catalan Agencia de Gestio
d'Ajuts Universitaris i de Recerca (AGAUR) with SGR 874
2014-2016, Fundaci\'o Privada Cellex Barcelona. 
N.S. was supported by the Erasmus
Mundus Doctorate Program Europhotonics (Grant No.
159224-1-2009-1-FR-ERA MUNDUS-EMJD). N.S., A.C.,
and M.L. acknowledge ERC AdG OSYRIS and EU FETPRO
QUIC. A. S. L. and L. O. acknowledge Max Planck Center for Attosecond Science (MPC-AS).


\end{document}